\documentclass[journal=jpclcd,preprint]{achemso}

\usepackage{pdfpages}
\usepackage{pgffor}

\usepackage{graphicx}%
\usepackage{dcolumn}%
\usepackage{bm}%

\usepackage{comment}
\usepackage[utf8]{inputenc}
\usepackage[T1]{fontenc}
\usepackage{mathptmx}
\usepackage{sansmath}
\usepackage{amsmath}
\usepackage{newtxtext,newtxmath}
\usepackage{physics}

\newcommand{\tensorsym}[1]{\boldsymbol{\mathsf{#1}}}
\newcommand{\der}[1]{\textrm{d}#1~}
\renewcommand{\tr}[1]{\textrm{Tr}\left[#1\right]}
\renewcommand{\exp}[1]{\textrm{exp}\left(#1\right)}
\newcommand{\Det}[1]{\textrm{Det}\left[#1\right]}

\newcommand{\qats}{ {\bf r} }
\newcommand{\qat}[1]{ r_{#1} }
\newcommand{\qatseq}{ {\bf r}^0 }

\newcommand{\fats}{ {\bf f}_{\qats} }

\newcommand{\qnms}{ {\bf q} }
\newcommand{\qnm}[1]{ q_{#1} }

\newcommand{\fnms}{ {\bf f}_{\qnms} }
\newcommand{\fnm}[1]{ {f_{\qnms}}_{#1} }

\newcommand{\pats}{ \nabla_{\qats} }
\newcommand{\pat}[1]{ \nabla_{\qat{#1}} }
\newcommand{\pnms}{ \nabla_{\qnms} }
\newcommand{\pnm}[1]{ \nabla_{\qnm{#1}} }

\newcommand{\enm}[1]{ E_{#1} }

\newcommand{\pot}{V}

\newcommand{\hessian}{\tensorsym{K}}
\newcommand{\massmat}{\tensorsym{M}}
\newcommand{\cellmat}{\tensorsym{h}}
\newcommand{\unimat}{\tensorsym{U}}
\newcommand{\qrmsmat}{\tensorsym{Q}}
\newcommand{\freqmat}{\boldsymbol{\Omega}}

\newcommand{\dispcorr}{\tensorsym{D}}

\newcommand{\vol}{\mathcal{V}}
\newcommand{\boltz}{k_B}
\newcommand{\ham}{\hat{H}}
\newcommand{\dimensionality}{d}

\newcommand{\com}{\textrm{cm}}
\newcommand{\har}{\textrm{har}}
\newcommand{\imf}{\textrm{imf}}
\newcommand{\vscf}{\textrm{vscf}}
\newcommand{\mptwo}{\textrm{mp2}}
\newcommand{\scp}{\textrm{scp}}

\newcommand{\trial}{\textrm{scp}}

\usepackage{xr}

\makeatletter
\newcommand*{\addFileDependency}[1]{%
  \typeout{(#1)}
  \@addtofilelist{#1}
  \IfFileExists{#1}{}{\typeout{No file #1.}}
}
\makeatother

\newcommand*{\myexternaldocument}[1]{%
    \externaldocument{#1}%
    \addFileDependency{#1.tex}%
    \addFileDependency{#1.aux}%
}
\usepackage[shortcuts]{extdash}

\myexternaldocument{si}

\title[]{An Assessment of Approximate Methods for Anharmonic Free Energies}

\author{Venkat Kapil}
\email{venkat.kapil@epfl.ch}
\altaffiliation{These authors contributed equally}
\affiliation{Laboratory of Computational Science and Modeling, Institut des Mat\'eriaux, \'Ecole Polytechnique F\'ed\'erale de Lausanne, 1015 Lausanne, Switzerland}
\author{Edgar Engel}
\email{edgar.engel@epfl.ch}
\altaffiliation{These authors contributed equally}
\affiliation{Laboratory of Computational Science and Modeling, Institut des Mat\'eriaux, \'Ecole Polytechnique F\'ed\'erale de Lausanne, 1015 Lausanne, Switzerland}
\author{Mariana Rossi}
\affiliation{Fritz Haber Institute of the Max Planck Society, Faradayweg 4-6, 14195 Berlin, Germany}
\author{Michele Ceriotti}
 \affiliation{Laboratory of Computational Science and Modeling, Institut des Mat\'eriaux, \'Ecole Polytechnique F\'ed\'erale de Lausanne, 1015 Lausanne, Switzerland}

\date{\today}

\begin{document}

\begin{tocentry}
\end{tocentry}

\begin{abstract}

Quantitative evaluations of the thermodynamic properties of materials -- most notably their stability, as measured by the free energy -- must take into account the role of thermal and zero-point energy fluctuations. 
While these effects can easily be estimated within a harmonic approximation, corrections arising from the anharmonic nature of the interatomic potential are often crucial and require computationally costly path integral simulations. %
Consequently, different approximate frameworks for computing affordable estimates of the anharmonic free energies have been developed over the years.
Understanding which of the approximations involved are justified for a given system, and therefore choosing the most suitable method, is complicated by the lack of comparative benchmarks. 
To facilitate this choice we assess the accuracy and efficiency of some of the most commonly used approximate methods -- the independent mode framework, the vibrational self-consistent field and self-consistent phonons -- by comparing the anharmonic correction to the Helmholtz free energy against reference path integral calculations.
These benchmarks are performed for a diverse set of systems, ranging from simple quasi-harmonic solids to flexible molecular crystals with freely-rotating units. Our results suggest that for simple solids such as allotropes of carbon these methods yield results that are in excellent agreement with the reference calculations, at a considerably lower computational cost. For more complex molecular systems such as polymorphs of ice and paracetamol the methods do not consistently provide a reliable approximation of the anharmonic correction. 
Despite substantial cancellation of errors 
when comparing the stability of different phases, we do not observe a systematic improvement over the harmonic approximation even for relative free-energies. Our results suggest that efforts towards obtaining computationally-feasible anharmonic free-energies for flexible molecular solids should therefore be directed towards reducing the expense of path integral methods. 
\end{abstract}
\maketitle

\section{\label{sec:introduction}Introduction}

The free energy is a key thermodynamic quantity which provides a measure of phase stability. Knowledge of the free energy and its derivatives with respect to temperature and applied fields can, in principle, be used to calculate every other thermodynamic observable. 
Reliable predictions of free energies from atomistic simulations remain a challenge because they require an accurate description of inter-atomic interactions, as well as proper treatment of the statistical mechanics of the nuclear degrees of freedom. 
The availability of computational resources combined with developments in electronic structure theory~\cite{lejaeghere_2016_dft, austin_2012_qmc, onida_2002_greensfunction, kotliar_2006_dmft, rohringer_2018_beyonddmft, neese_2019_cc, riplinger_2016_lscc, Gruber_2018_pcc, caldeweyher_2018_mcdft, zen_2018_dmc} have made it possible to calculate the Born-Oppenheimer (BO) surfaces that govern nuclear motion routinely and accurately. 
Consequently, the accuracy of free energy calculations is often limited by the statistical sampling of the nuclear degrees of freedom~\cite{hoja_2019_molcrys}. 
This is most commonly performed within a harmonic approximation, which is reasonable for quasi-harmonic systems, such as metals at low temperatures~\cite{grabowski_2009_al}, but fails close to the melting temperature~\cite{grimvall_2012_metals,grabowski_2009_al} and in the presence of defects~\cite{glensk_2014_defects}.
The problem is exacerbated for the case of organic solids, which require a proper description of anharmonicity arising from quantum nuclear motion even at room temperature~\cite{li_2011_hb,fang_2016_bio,rossi_2015_bio, rossi_2016_paracetamol}. 
This is similarly true for (i) systems containing light elements, such as hydrogen~\cite{azadi_2014_hydrogen,bianco_2018_hydrogen, fang_2019_hydrogen}, helium~\cite{ceperley_1995_helium} and water~\cite{ceriotti_2016_water}, ice~\cite{ramirez_2010_ice,engel_2015_ice,cheng_2019_ice}, metal organic frameworks ~\cite{lamaire_2019_mofs, kapil_2019_mofs}, the record high-$T_c$ conventional superconductor SH$_3$~\cite{errea_2015_sh3}, (ii) systems of reduced effective dimensionality such as graphene~\cite{lin_2011_graphene,ramire_2016_graphene} and (iii) molecular systems such as paracetamol~\cite{rossi_2016_paracetamol}. 

Within the BO approximation\cite{born_1927_bo_approximation} and given the BO potential, exact anharmonic free energies can be calculated using approaches based on imaginary time path integral (PI) simulations~\cite{chandler_1981_pi,parrinello_1984}. However, the required number of force evaluations, when combined with accurate electronic structure calculations, render these approaches impractical for any but the smallest systems. Consequently, a small zoo of frameworks has been developed which approximately account for quantum anharmonic motion at a much lower computational cost. These invoke different approximations and exhibit different scaling behaviour with system size. 

In this work, we present an extensive benchmark of the accuracy of some of the most common approximate techniques — such as the harmonic approximation (HAR)~\cite{ashcroft_1976_lattice_vibrations}, self-consistent phonons (SCP)~\cite{brown_2013_scp,errea_2013_scp}, the independent mode framework (IMF)~\cite{monserrat_2013_vscf}, and the vibrational self-consistent field (VSCF)~\cite{monserrat_2013_vscf}  — against reference results obtained using PI thermodynamic integration (QTI)~\cite{mcbride_2012_qti, rossi_2016_paracetamol}. Computationally efficient algorithms for these methods have been developed and implemented in the universal force engine i-PI \cite{kapil_2018_ipi}.  
The accuracy and the computational efficiency of the methods is tested on a set of solids ranging from from simple allotropes of carbon, anharmonic but relatively rigid polymorphs of ice to polymorphs of paracetamol that contain (nearly) freely rotating internal degrees of freedom. %

\section{\label{sec:theory}Theory}

To briefly outline the different free energy methods we discuss, we consider a three-dimensional periodic system, whose minimum potential energy, ``equilibrium'' atomic positions form a Bravais lattice, noting that finite and aperiodic systems simply represent the limit of infinite period. 
The full ionic Hamiltonian of such a system is 
\begin{equation}
	\ham = - \sum_{p,i} \frac{\hbar^2}{2m_i} \nabla_{{\bf r}_{pi}}^2 + V \left(\left\{ {\bf r}_{pi} \right\} \right)
	\label{eq:full_ionic_hamiltonian}
\end{equation}
where $m_i$ is the mass of nucleus $i$, $V$ is the BO potential governing nuclear motion, and $p$ and $i$ run over the Bravais points and the nuclei within a unit cell, respectively. In practice we perform supercell simulations using periodic Born-von\,Karman simulation cells $\cellmat$ consisting of $N_a \times N_b \times N_c$ replicas of the unit cell and with cell vectors $N_a {\bf R}_a$, $N_b {\bf R}_b$, and $N_c {\bf R}_c$.

In the following we only consider $\Gamma$-point vibrational motion within the simulation cell. We thereby sample those ${\bf K}$-points within the first vibrational Brillouin Zone (BZ) of the underlying unit cell, for which $\exp{-i {\bf K} \cdot {\bf R}} = 1 \quad \forall \quad {\bf R} = N_a {\bf R}_a + N_b {\bf R}_b + N_c {\bf R}_c$. The Hamiltonian of the system is then uniquely defined given the positions of the $N$ particles within the simulation cell $\cellmat$:
\begin{equation}
    \ham = - \frac{\hbar^2}{2} \pats \massmat^{-1} \pats^{T}  + \pot\left( \qats, \cellmat \right)
    \label{eq:ionic_hamiltonian}    
\end{equation}
where $(\pats, \qats) \equiv (\{ \pat{1}, \dots, \pat{3N} \}, \{ \qat{1}, \dots, \qat{3N} \})$ denotes the momenta and positions of the $3N$ degrees of freedom associated with the $N$ particles and $\massmat = \text{Diag}\left[m_1, \dots, m_{3N}\right]$. The canonical partition function of the system at inverse temperature $\beta =(\boltz T)^{-1}$ and volume $\vol = \Det{\cellmat}$ is defined as
\begin{equation}
    Z(N,\vol, \beta) = \tr{\exp{-\beta \ham}} \,, 
    \label{eq:partition_function}    
\end{equation}
where the trace can be performed over any complete basis set. In the thermodynamic limit, the Helmholtz free energy of the system is
\begin{equation}
        A(N,\vol, \beta) 
        = - \beta^{-1} \ln Z(N,\vol, \beta) \,.
    \label{eq:free_energy}    
\end{equation}
Direct computation of $A$ is hindered by the computational complexity of solving the Schr\"odinger equation associated with the Hamiltonian $\ham$, motivating approximate but computationally more affordable approaches. 

\subsection{\label{subsec:harmonic}Harmonic Approximation}
For small displacements, $\qats - \qatseq$, of the particles from their equilibrium positions, $\qatseq \equiv \textrm{argmin}_{\qats}~V(\qats,\cellmat)$, the potential can be Taylor expanded. Truncation after the quadratic term amounts to the harmonic approximation
\begin{equation}
    V^{\har}\left( \qats, \cellmat \right) 
    = V^{(0)} + \frac{1}{2} ( \qats - \qatseq ) \hessian
    ( \qats - \qatseq )^{T}
    \label{eq:har_potential}    
\end{equation}
with $V^{(0)} \equiv V(\qatseq)$ and $\hessian = \left.\nabla^{2} V(\qats,\cellmat)\right|_{\qats = \qatseq}$. The spectral decomposition of the Hessian, 
\begin{equation}
    \hessian = \massmat^{\frac{1}{2}} \unimat \freqmat^2 \unimat^{T}  \massmat^{\frac{1}{2}} = \tilde{\unimat} \freqmat^2  \tilde{\unimat}^{T} 
    \label{eq:hessian_diagonalisation}
\end{equation}
provides the unitary matrix $\unimat$, the mass-scaled transformation matrix $\tilde{\unimat}$, and the diagonal matrix containing the normal mode frequencies $\freqmat = \text{Diag}\left[\omega_1, \dots, \omega_{3N}\right]$.
After transformation to the normal mode coordinates $\pnms \equiv \tilde{\unimat}^{T} \pats$ and $\qnms \equiv \tilde{\unimat} (\qats - \qatseq)$, the Hamiltonian 
\begin{equation}
    \begin{split}
        \ham^{\har} 
        & = - \frac{\hbar^2}{2} \pnms^{2} +  \frac{1}{2} \qnms \freqmat^2 \qnms^{T} + V(\qatseq) \\ 
        & = V^{(0)} + \ham^{\com}  + \sum_{i=1}^{\dimensionality} \left[ - \frac{\hbar^2}{2} \pnm{i}^{2} +  \frac{1}{2} \omega_i^2 \qnm{i}^2 \right]
    \end{split}
    \label{eq:har_hamiltonian}
\end{equation}
separates into $\pot^{(0)}$, the centre of mass term $\ham^{\com}$, and a term describing a system of $\dimensionality=3N-3$ independent simple harmonic oscillators (SHO) whose energies and wave functions for a given excitation state $s_i$, $\enm{i,(s_i)}^{\har} = (s_i + 1/2) \hbar \omega_{i}$ and $\vert \phi^{(s_i)}_{i} \rangle$, are known analytically. In finite systems, global rotations decouple analogously. The centre of mass contribution to the free energy $A^{\com}$ is that of a free particle in a three dimensional box with a volume and shape equal to that of the Wigner-Seitz cell of the system, while the contribution from the free rotations of finite systems can be computed within the rigid rotor approximation\cite{mcquarrie_1997_rigidrotor}. \\

For the remaining system of harmonic oscillators, the $\dimensionality$-body wave function of the global state described by the $\dimensionality$-tuple ${\bf s} = \left(s_{1}, \dots, s_{\dimensionality}\right)$ is a Hartree product of the independent normal mode wave functions:
\begin{equation}
    \vert \Psi^{\har}_{\left(\mathbf{s}\right)} \rangle = \prod_{i=1}^{\dimensionality} \otimes \ket{\phi_{i,(s_{i})}}
    \label{eq:har_wavefunction}
\end{equation}
and the free energy is:
\begin{equation}
   A^{\har}(N,\vol,\beta)
   = V^{(0)} + A^{\com} + \sum_{i=1}^{d} 
   \left[ 
   \frac{\hbar \omega_{i}}{2} 
   + \beta^{-1} \ln 
   \left(
   1 - e^{-\beta \hbar \omega_{i}}
   \right) 
   \right].
   \label{eq:har_free_energy}
\end{equation}
\subsection{\label{subsec:imf}Independent Mode Framework}

A first approximation to anharmonic quantum nuclear motion is detailed in the work of Monserrat \textit{et al.}\,\cite{monserrat_2013_vscf}. The potential is expanded in terms of the normal mode coordinates
\begin{equation}
	V(\qnms) = V^0 + \sum_{i}^{d} V^{(1)}(\qnm{i})
	+ \frac{1}{2} \sum_{i}^{d} \sum_{j \neq i}^{d} 
	V^{(2)}(\qnm{i},\qnm{j}) + \cdots \,,
    \label{eq:potential_expansion}
\end{equation}
where
\begin{equation}
	V^{(1)}(\qnm{i}) = 
	V(0,\ldots,\qnm{i},\ldots,0) - V^{(0)} \, ,
\label{eq:indep_mode_potential}
\end{equation}
is the (anharmonic) independent mode term and 
\begin{equation}
\begin{split}
   V^{(2)}(\qnm{i},\qnm{j}) = 
   & V(0,\ldots,\qnm{i},\ldots,\qnm{j},\ldots,0) \\
   & - V^{(1)}(\qnm{i}) - V^{(1)}(\qnm{j}) - V^{(0)} \, .
\end{split}
\label{eq:coupled_mode_potential}
\end{equation}
describes pairwise coupling between normal modes. This expansion can be continued for more general $n$-body terms $V^{(n)}$. Since one starts with the harmonic approximation, in which the normal modes are non-interacting, the hope is that higher-order terms decrease in size with increasing $n$. The validity of this assumption is discussed in section~\ref{sec:rnd}. Truncation after $V^{(1)}$ amounts to the independent mode approximation with the Hamiltonian 
\begin{equation}
	\ham^{\imf} = V^{(0)} + \sum_{i}^{d} \left[ - \frac{\hbar^2}{2}\pnm{i}^2 + V^{(1)}(\qnm{i}) \right]\,.
	\label{eq:imf_hamiltonian}
\end{equation}
Despite the presence of anharmonicity the normal modes remain independent and a Hartree product analogous to Eq.~(\ref{eq:har_wavefunction})
of anharmonic normal mode wave functions solves the Schr{\"o}dinger equation yielding the eigenvalues $E^{\imf}_{i,(s_{i})}$. The Helmholtz free energy is 
\begin{equation}
   A^{\imf}(N,\vol,\beta) 
   = V^{(0)} + A^{\com} - \sum_{i=1}^{d} 
   \left[ \beta^{-1}  \ln{\sum_{s_{i}} \exp{-\beta E^{\imf}_{i,(s_{i})}} }
   \right].
   \label{eq:imf_free_energy}
\end{equation}

\subsection{\label{subsec:vscf}Vibrational Self-consistent Field}

Retaining terms involving $V^{(2)}$ (and / or higher order terms) leads to coupling of the previously independent normal modes and complicates the solution of the Schr\"odinger equation. Monserrat \textit{et al.}\,\cite{monserrat_2013_vscf} solve the equation 
\begin{equation}
	\left[ - \frac{\hbar^2}{2} \pnms^2 + V(\qnms) \right] \ket{ \Psi^{\vscf}_{\left(\mathbf{s}\right)}} = E^{\vscf}_{\left(\mathbf{s}\right)} \ket{ \Psi^{\vscf}_{\left(\mathbf{s}\right)}}
	\label{eq:vscf_schroedinger}
\end{equation}
within the iterative Vibrational Self-Consistent Field (VSCF) approach, where $V(\qnms)$ represents the truncated form of Eq.~\eqref{eq:potential_expansion}. Using a Hartree product trial wavefunction amounts to a mean-field (MF) treatment  and leads to the VSCF equations
\begin{equation}
	\left[
		- \frac{\hbar^2}{2} \pnm{i}^2 + {\bar V}_{i}(\qnm{i})
	\right]
	\ket{\psi^{\vscf}_{i,(s_{i})}}  =
	E^{\vscf}_{i} \ket{\psi^{\vscf}_{i,(s_{i})}}
	\label{eq:MF_schroedinger}
\end{equation}
where ${\bar V}_{i}(\qnm{i})$ is the mean-field potential experienced by normal mode $i$,
\begin{equation}
	{\bar V}_{i}(\qnm{i}) =
	\sum_{j \neq i} \rho(\qnm{j})~ V(\qnms)
	\label{eq:MF_potential}
\end{equation}
with
\begin{equation}
\rho(\qnm{j}) = 
	\frac{\sum_{s_j}
	\exp{ - \beta E^{\vscf}_{j,(s_j)} }
	\ket{\psi^{\vscf}_{j,(s_j)}} \bra{\psi^{\vscf}_{j,(s_j)}}}
    {\sum_{s_j}
	\exp{ - \beta E^{\vscf}_{j,(s_j)} }}\,.
\end{equation}
To lowest order the VSCF free energy becomes 
\begin{equation}
    A^{\vscf}(N,\vol,\beta)  = V^{(0)} + A^{\com} - \beta^{-1} \ln \sum_{{\bf s}} \exp{ - \beta \sum_{i} E^{\vscf}_{i,(s_i)}}\,.
\end{equation}
A perturbation theory can be constructed in terms of the (assumed to be small) difference between the mapped out and the MF potential, $V(\qnms) - \sum_{i} {\bar V}_{i}(\qnm{i})$, leading to a second-order MP2 correction to the energy of state ${\bf s}$ given by
\begin{equation}
	E_{\left(\mathbf{s}\right)}^{\vscf,(2)} = \sum_{{\bf s}' \neq {\bf s}} 
	\frac {\mel{\Psi^{\vscf}_{\left(\mathbf{s}'\right)} }{V(\qnms) - \sum_{i} {\bar V}_{i}(\qnm{i})}{\Psi^{\vscf}_{\left(\mathbf{s}\right)} }^{2}}
	{E_{\bf s}^{\vscf,(1)} - E_{{\bf s}'}^{\vscf,(1)}}
	 \label{eq:mptwo_correction}
\end{equation}
and the approximate free energy
\begin{equation}
    \begin{split}
	    A^{\mptwo}(N,\vol,\beta)  = A^{\vscf}(N,\vol,\beta) - \beta^{-1} \ln \sum_{{\bf s}} \exp{ - \beta E^{\vscf,(2)}_{{\bf s}}}
     \end{split}
    \label{eq:vscf_free_energy}
\end{equation}

\subsection{\label{subsec:scp}Self-consistent Phonons}

Another way of calculating an anharmonic correction on top of a harmonic approximation exploits the Gibbs-Bogoliubov inequality\cite{callen_1985_bogoliubov}, which states that the true free energy of a system is always bounded from above by the free energy $A^{\scp}$ computed using a trial density matrix, $\hat{\rho}^{\trial}$: 
\begin{equation}
  A  < A^{\scp} = \expectationvalue{\hat{H} + \beta^{-1} \ln \hat{\rho}^{\trial}}_{\hat{H}^{\trial}} \quad ; \quad \hat{\rho}^{\trial} = \frac{\exp{-\beta \hat{H}^{\trial}}}{\tr{\exp{-\beta \hat{H}^{\trial}}}}
\end{equation}
where $\expectationvalue{\square}_{\hat{H}^{\trial}} = \tr{\rho^{\trial} ~ \square}$ is an ensemble average defined by the the trial density matrix $\hat{\rho}^{\trial}$. Within the self-consistent phonons method\cite{hooton_1955_scp,hooton_1955_scp2}, $\hat{\rho}^{\trial}$ is the density matrix of a harmonic Hamiltonian with Hessian $\hessian^{\trial}$ and equilibrium positions $\qats^{\trial}$:
\begin{equation}
    \hat{\rho}^{\trial}(\qats) 
    = (2\pi\dispcorr)^{-\frac{1}{2}} \exp{-\frac{1}{2}(\qats - \qats^{\trial}) \dispcorr^{-1} (\qats - \qats^{\trial})^{T}}
\end{equation}
where, $\dispcorr = \massmat^{-\frac{1}{2}} \unimat  \qrmsmat^{2} \unimat^{T}\massmat^{-\frac{1}{2}}$ and $\qrmsmat$ is a diagonal matrix containing the root-mean-square (RMS) displacements~\cite{brown_2013_scp}:
\begin{equation}
    \tilde{\qnm{i}}^{\trial}(T) = \sqrt{\frac{\hbar}{2 \omega^{\trial}_{i}} \coth{\frac{\hbar \omega^{\trial}_{i}}{2 \boltz T} }}
\end{equation}
of the normal modes. The lowest upper bound to the true free energy is obtained by minimizing the free energy with respect to $\qats^{\trial}$ and $\hessian^{\trial}$. This leads to the steady state conditions\cite{brown_2013_scp}
\begin{equation}
\begin{split}
    \expectationvalue{\fats(\qats)}_{H^{\trial}} & = 0  \\
    \expectationvalue{\hessian(\qats)}_{H^{\trial}} & = \hessian^{\trial}
\end{split}
\end{equation}
where $\fats(\qats)$ correspond to the forces of the potential $\pot ({\qats})$. The solution is obtained in a self consistent manner by starting with educated guesses of $\left(\qats^{\trial}, \hessian^{\trial}\right)  = \left(\qats^{\trial}_{0}, \hessian^{\trial}_{0}\right)$ -- which are in practice chosen to be those obtained within the harmonic approximation -- and updating 
\begin{equation}
\begin{split}
 \hessian^{\trial}_{\ham^{\trial}_{l+1}} &= \expectationvalue{\hessian(\qats)}_{\ham^{\trial}_{l}} \\
 \qats^{\trial}_{\ham^{\trial}_{l+1}} &= \qats^{\trial}_{\ham^{\trial}_{l}} + {\hessian^{\trial}_{\ham^{\trial}_{l+1}}}^{-1}   \expectationvalue{\fats}_{\ham^{\trial}_{l}}
 \label{eq:scp_optimization}
\end{split}
\end{equation}
until convergence is achieved. Here $\ham^{\trial}_l$ denotes the the trial Hamiltonian of the $l$-th SCP iteration. The resultant free energy at the $l$-th iteration is  calculated as:
\begin{equation}
\begin{split}
    A^{\scp} &=  A^{\com} +    \left[ 
   \frac{\hbar \omega^{\trial,l}_{i}}{2} 
   + \beta^{-1} \ln 
   \left(
   1 - e^{-\beta \hbar \omega^{\trial,l}_{i}}
   \right) 
   \right]  \\ 
             & + \expectationvalue{V(\qats) - \frac{1}{2} ( \qats - \qats^{\trial}_{l} ) \hessian^{\trial}_{l} ( \qats - \qats^{\trial}_{l} )^{T}}_{\ham^{\trial}_{l}}\,.
\end{split}
\end{equation}

\subsection{\label{subsec:ti}Thermodynamic Integration}

Within the thermodynamic integration scheme the free energy differences between two states is calculated as the work to reversibly transform one state into the other~\cite{kirkwood_1935_ti, ghiringhelli_2005_ti,cheng_2018_ti_paths}. 
For solids this method can be used to calculate the classical anharmonic correction to the harmonic Helmholtz free energy as the reversible work done while ``switching on" the anharmonic part of the potential~\cite{frenkel_1984_cti,habershon_2011_cti_h2o}. 
In the Hamiltonian $H^{\lambda} = (1- \lambda) H^{\har} + \lambda~ H$ the Kirkwoord coupling parameter $\lambda$ smoothly switches the potential from harmonic $(\lambda=0)$ to fully anharmonic $(\lambda=1)$.
The free energy difference is obtained by computing the integral of the thermodynamic force along the switching path: 
\begin{equation}
    \Delta A_{\textrm{cl}} =  A_{\textrm{cl}} - A_{\textrm{cl}}^{\har} 
    = \int_{0}^{1} \der{\lambda} \left(\frac{\partial A}{\partial \lambda}\right) = \int_{0}^{1} \textrm{d}\lambda \expectationvalue{ V - V^{\har}}_{H^{\lambda}} ,
\label{eq:ti_classical}
\end{equation}
where, $\expectationvalue{\square}_{H^{\lambda}}$ represents an average over the classical canonical ensemble sampled by the intermediate Hamiltonian and $A_{\textrm{cl}}^{\har}$ is the classical harmonic free energy. Setting aside statistical errors, $\Delta A_{\textrm{cl}}$ can be computed exactly by sampling the thermodynamic forces at multiple values of $\lambda \in [0,1]$ using molecular dynamics simulations. Alternatively, $\Delta A_{\textrm{cl}}$ can also be calculated by performing a thermodynamic integration from a low temperature harmonic solid~\cite{hoover_1971_cti, moustafa_2015_ti} as 
\begin{equation}
    \Delta A_{\textrm{cl}} =  - T \int_{0}^{T} \der{\tilde{T}} \frac{\expectationvalue{ V - V^{(0)} - \frac{3N}{2} k_B \tilde{T}}_{\tilde{T}}}{{\tilde{T}}^2},
\label{eq:ti_classical_2}
\end{equation}
where $\expectationvalue{\square}_{\tilde{T}}$ is an average over the $N\mathcal{V} \tilde{T}$ ensemble. 

To include quantum anharmonic corrections due to zero- point energy, tunnelling, etc., a second thermodynamic integration must be performed to calculate the work required to reversibly transform the particles from classical to quantum~\cite{morales_1991_qti_cl2qn, habershon_2011_qti_cl2qn}. This can be achieved by defining the Hamiltonian $\hat{H}^{g} = - g \frac{\hbar^2}{2} \pats \massmat^{-1} \pats^{T}  + \pot\left( \qats, \cellmat \right) $, where $g$ scales the mass of the particles\cite{mcbride_2012_qti, fang_2016_bio, ceriotti_2013_isofrac}. As $g$ is varied from 1 to 0, the de~Broglie wavelength of the particles smoothly drops from its physical value to zero, yielding the desired transformation from quantum to classical particles. The corresponding free energy difference is:
\begin{equation}
    \Delta A_{\textrm{qn}} = A_{\textrm{qn}} - A_{\textrm{cl}} = \int_{0}^{1} \der{g} g^{-1} \expectationvalue{\hat{T} - T_{\textrm{cl}}}_{\hat{H}^{g}}.
    \label{eq:ti_quantum}
\end{equation}
where $\expectationvalue{\hat{T}}_{\hat{H}^{g}}$ represents the average kinetic energy for the intermediate Hamiltonian and $T_{\textrm{cl}}$ is the classical kinetic energy which is independent of the mass of the system. Eq.~(\ref{eq:ti_quantum}) can be computed exactly (modulo statistical error) by sampling the quantum canonical ensembles for $g \in [0,1]$ using PI molecular dynamics (PIMD). The difference between the classical and quantum kinetic energy can be computed directly using a centroid-virial kinetic energy estimator~\cite{kolr_1996_cvest, ceperley_1995_helium}. The total anharmonic free energy is computed as:
\begin{equation}
    A =   A^{\com} + A_{\textrm{cl}}^{\har} + \Delta A_{\textrm{cl}} + \Delta A_{\textrm{qn}}
\end{equation}

\section{\label{sec:implementation}Implementation}

\begin{figure}
    \centering
    \includegraphics[width=0.50\textwidth]{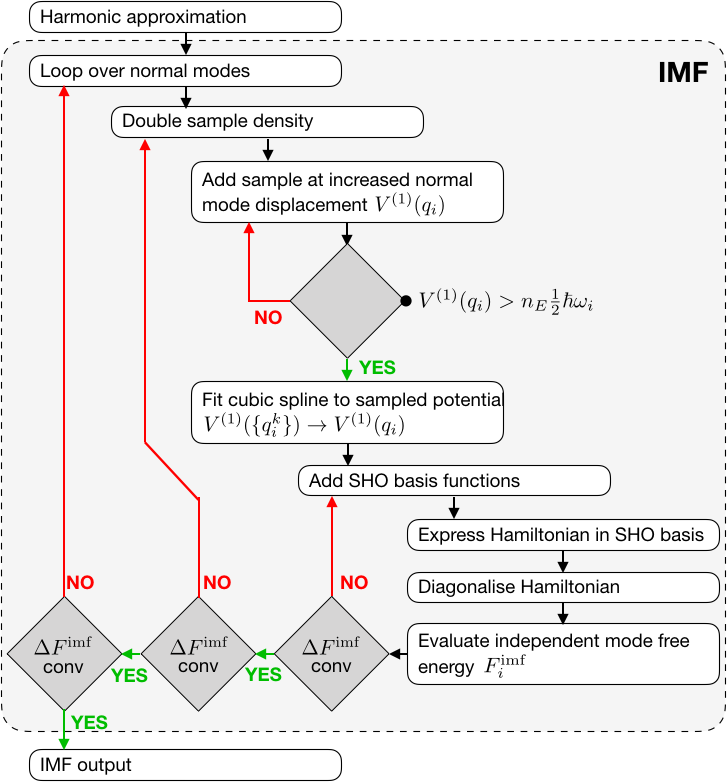}
    \caption{\label{fig:imf_scheme}Schematic representation of the independent mode approximation module.}
\end{figure}

\begin{figure}
    \centering
    (a)\phantom{aaaaaaaaaaaaaaaaaaaaaaaaaaaaaaaaaaaaaaaaaaaaaaaaaaaaaaaaaaaa}\\
    \includegraphics[width=0.5\textwidth]{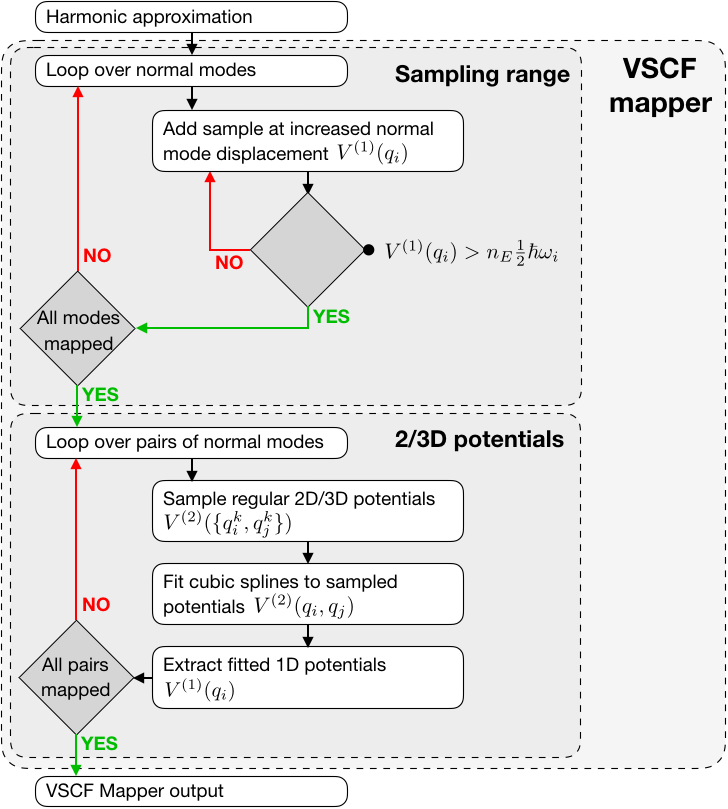}
    (b)\phantom{aaaaaaaaaaaaaaaaaaaaaaaaaaaaaaaaaaaaaaaaaaaaaaaaaaaaaaaaaaaa}\\
    \includegraphics[width=0.5\textwidth]{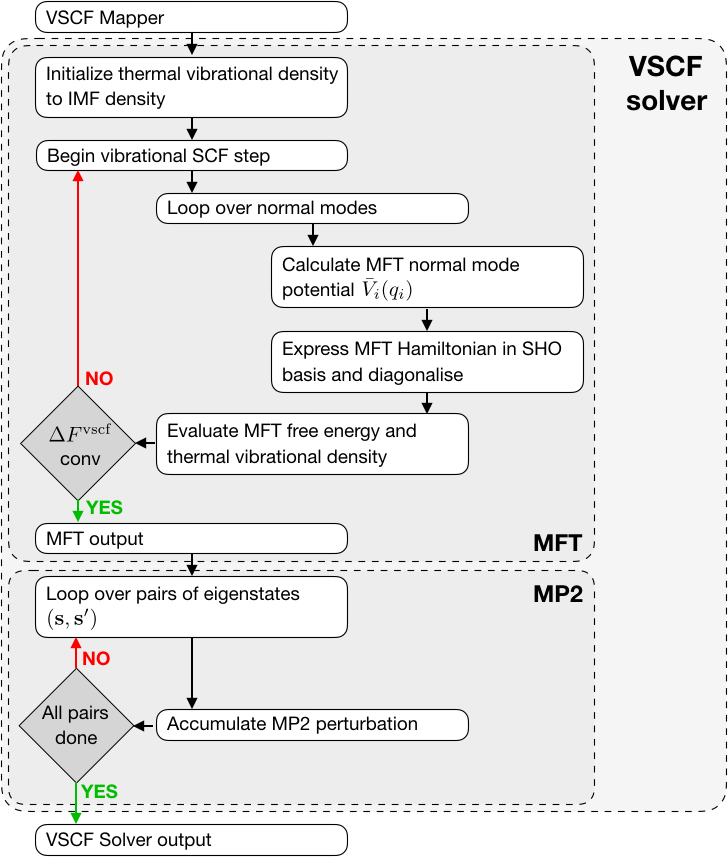}
    \caption{\label{fig:vscf_scheme}Schematic representation of the (a) VSCF mapper and (b) VSCF solver module.}
\end{figure}

In order to perform a direct comparison of the different approximate methods, IMF, VSCF, and SCP were implemented within i-PI~\cite{kapil_2018_ipi}, an open-source python package for atomistic simulations, which couples to a variety of density-functional-theory (DFT) and empirical and machine-learning potential codes. 
The IMF, VSCF, and SCP implementations are schematically shown in Figs.~\ref{fig:imf_scheme} to \ref{fig:scp_scheme}. The reference free energies can simply be evaluated on the basis of (PI)MD simulations and require no dedicated implementation.

\subsection{\label{subsec:impl:imf}Independent Mode Framework}

Given the transformation matrix, ${\tilde {\bf U}}$, and normal-mode frequencies from the harmonic approximation, we perform single-point energy and force evaluations for equally spaced configurations $\qnm{i}^j = j f \tilde{\qnm{i}}(T)$ along each normal mode $i$ where $\tilde{\qnm{i}}$ is the RMS displacement of the normal mode at a target temperature $T$.  We increase $j$ by one at a time until the sampled energy $V^{(1)}(\qnm{i}^j)$ exceeds a user-defined multiple $n_E$ of the thermal harmonic energy $V^{(1)}(\qnm{i}^j) > n_E E^{\har}_{i}(T)$.
This ensures that the potential is always mapped out far enough into the classically forbidden region (but only as far as necessary) to localize the nuclear density, at temperatures lower or equal than the chosen target.
The independent mode potential $\sum_i V^{(1)}(\qnm{i})$ is then reconstructed by fitting cubic splines to $\{ (\qnm{i}^j, V^{(1)}(\qnm{i}^j)) \}$. 
The corresponding independent mode Hamiltonian is expanded in a basis of SHO eigenstates and diagonalized to evaluate the independent mode anharmonic Helmholtz free energy. 
The Helmholtz free energy is converged with respect to the density of the frozen-phonon samples $\qnm{i}^j$ by repeatedly halving $f$ and supplementing the already collected $\{ (\qnm{i}^j, V^{(1)}(\qnm{i}^j)) \}$ with corresponding samples, until the required convergence threshold is met.
For each $f$ the Helmholtz free energy is converged with respect to the size of the SHO basis. 

\subsection{\label{subsec:impl:vscf}Vibrational Self-Consistent Field}

The implementation of the VSCF framework is split into two modules: one for mapping the potential energy surface (PES), and one for solving the VSCF problem.
The mapping strategy mirrors that employed in the IMF module. In a first loop over normal modes, we collect $\{ (j f \tilde{\qnm{i}}(T), V^{(1)}(j f \tilde{\qnm{i}}(T)) \}$ until the sampled potential exceeds a user-defined multiple of the harmonic energy $V^{(1)}(j f \tilde{\qnm{i}}(T)) > n_E E^{\har}_{i}(T)$, thereby also determining the sampling range for the coupling corrections.
In a second loop over $n$-tuples of normal modes, we then sample $\{ ((\qnm{i_1}^{j_1},\ldots,\qnm{i_n}^{j_n}), V^{(n)}(\qnm{i_1}^{j_1},\ldots,\qnm{i_n}^{j_n})) \}$ in a similar fashion and extract the coupling corrections $V^{(n)}(\qnm{i_1},\ldots,\qnm{i_n})$ using cubic spline fits. 
Currently, sampling and fitting of $n = 2,3$ are implemented.
The extracted coupling corrections are stored for use within the VSCF solver module.

The module for solving the VSCF problem consists of two submodules, the first of which performs the VSCF calculation itself.
The thermal density determining the mean-field potentials $\{ {\bar V}_i(\qnm{i}) \}$ is initialised as the IMF thermal density.
Within a VSCF step the MF independent mode Hamiltonians for the given MF potentials are constructed, expanded in an SHO basis, and diagonalized to determine the updated MF thermal vibrational density and the free energy $A^{\vscf}$. To stabilize the VSCF convergence, 50\% of the thermal density resulting from the previous VSCF iteration are mixed in before the mean-field potentials $\{ {\bar V}_i(\qnm{i}) \}$ are updated and the next VSCF step is initiated.
This is repeated until self-consistency has been reached as indicated by convergence of the associated free energy $A^{\vscf}$ to within the required threshold.

The second submodule allows the calculation of an MP2 correction on top of the MF eigenstates and -energies by looping over pairs of eigenstates $({\bf s}, {\bf s}')$ to evaluate the MP2 corrections in Eq.~(\ref{eq:mptwo_correction}) on a real-space grid of predefined density. Care is taken to only consider eigenstates of the self-consistent MF description with eigenenergies $E^{\vscf}_{{\bf s}}$ within a set multiple of $\boltz T$.

\subsection{\label{subsec:impl:scp}Self-Consistent Phonons}
\begin{figure}
    \centering
    \includegraphics[width=0.50\textwidth]{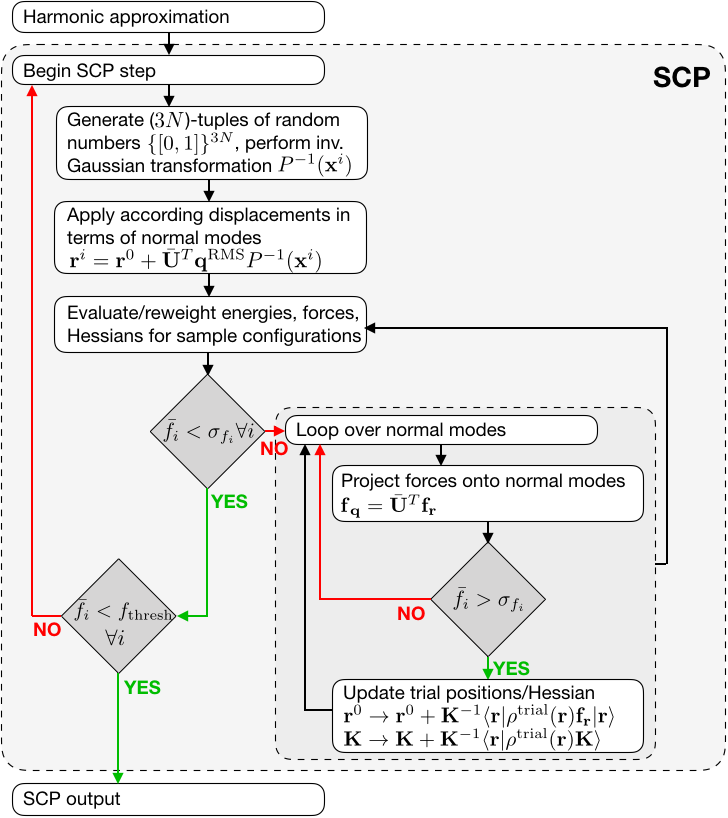}
    \caption{\label{fig:scp_scheme}Schematic representation of the SCP module.}
\end{figure}

Our implementation of the SCP method is schematically shown in Fig.~\ref{fig:scp_scheme}. 
In the first loop over SCP steps, we construct the trial density matrix $\hat{\rho}^{\trial}(\qats)$ using the mean position $\qats^{\trial}$ and Hessian $\hessian^{\trial}$ obtained from the previous step. In the first step $\hat{\rho}^{\trial}(\qats)$ are those obtained within the harmonic approximation.
In the second loop we calculate the ensemble averages of the forces and the Hessian, necessary to perform the optimization steps described in equation \ref{eq:scp_optimization}. These are realized as Gaussian integrals and computed using MC importance sampling as
\begin{equation}
 \expectationvalue{\square}_{H^{\trial}} = \frac{1}{N_s} \sum_{i=1}^{N_s} \square + \mathcal{O} \left( 1/\sqrt{N_s} \right)
 \label{eq:mc_error}
\end{equation}
where $N_s$ is the number of samples. 
Samples are generated by translating $3(N-1)$-tuples of (quasi-)random numbers on the interval $[0,1]$ into atomic displacements from the mean position $\qats^{\trial}$ using the inverse cumulative distribution function of $\rho^{\trial}(\qats)$ with a Beasley-Springer-Moro algorithm~\cite{beasley_1977_inv_gaussian, moro_1995_inv_gaussian}.
To speed up the convergence of the averages with respect to the number of samples, we employ the following tricks:
\begin{enumerate}
    \item 
    For  small system sizes, instead of drawing pseudo random numbers, we use low-discrepancy quasi-random numbers -- specifically Sobol sequences~\cite{sobol_1977_sequence} -- as was done in the implementation of Brown and coworkers~\cite{brown_2013_scp}. This leads to a more uniform sampling, so that error in equation \ref{eq:mc_error} decays as $\mathcal{O} \left( \ln(N_s)^d/N_s \right)$ which becomes $\sim  \mathcal{O} \left(1 /N\right)$ for low dimensional integrals\cite{schrer_2003_sobol_efficiency}. For large system sizes, we resort to pseudo random numbers as the performance of Sobol sequences degrades\cite{lemieux_2009_qmc}.  We use the FORTRAN implementation of Burkardt~\cite{burkardt_sobol} to generate Sobol sequences. 
    
    \item 
    As was done in the implementation of Errea and co-workers\cite{errea_2014_scp},  we re-use samples from previous SCP iterations via a reweighting scheme. Given the updated trial density $\hat{\rho}^{\trial}_{l}$ at the $l$-th SCF iteration, the reweighted average using the $N_s$ samples $\{ \qats \}^k$ drawn from the trial density $\hat{\rho}^{\trial}_{k}$ at the $k$-th SCP iteration is
    \begin{equation}
    \begin{split}
        \expectationvalue{\square}_{\ham^{\trial}_{l}}^{k} 
        & = \frac{1}{N_s} \sum_{\qats \in \{ \qats \}^k} \left[ w_l^k(\qats) \square (\qats) \right]
        \\
        w_l^k(\qats) & 
        \equiv \rho^{\trial}_{l}(\qats) / \rho^{\trial}_{k}(\qats)
    \end{split}
    \label{eq:reweighting}
    \end{equation}
    We minimize the error in the global estimates $\expectationvalue{\square}_l$ at the $l$-th SCP iteration by weighting samples drawn in the $k$-th SCP iteration according to a ``batch weight'' 
    \begin{equation}
    \begin{split}
         \expectationvalue{\square}_{\ham^{\trial}_{l}}
         &\approx \sum_{k=1}^{l}  \left[ W^k \expectationvalue{\square}_l^{k} \right] \\
         W^k &= \textrm{Var}\left( \square^k \right)^{-1} \left[ \sum_{k'} \textrm{Var}\left( \square^{k'} \right)^{-1} \right]^{-1}
    \end{split}
    \end{equation}
    where $\square^k \equiv \square (\{\qats\}^k)$ and the variance of a generic observable over samples from $k$-th SCP iteration is~\cite{ceriotti_2011_curse}
    \begin{equation}
        \textrm{Var}\left( \square^k \right) = 
        \left( \textrm{Var}\left( \square \right) + \expectationvalue{-\square \ln{w_l^k}} \right) \frac{\exp{\textrm{Var}\left( -\ln{w_l^k} \right)}}{N_s}
    \end{equation}
    where $w_l^k \equiv w_l^k\left( \{ \qats \}^k \right)$, provided both $\square$ and $-\ln{w_l^k}$ are normally distributed.
    Neglecting $\expectationvalue{-\square \ln{w_l^k}}$ renders the batch weights independent of the observable being considered
    \begin{equation}
        W_l^k = \frac{\exp{-\textrm{Var}\left( \ln{w_l^k \left( \{ \qats \}^k \right)} \right)}}{\sum_{m} \exp{-\textrm{Var}\left( \ln{w_l^m \left( \{ \qats \}^m \right)} \right)}}
    \end{equation}
    and thereby suitable for both Hessians and forces.

    \item 
    Taking inspiration from stochastic over-relaxation algorithms\cite{alder_1981_overrelaxation, ceriotti_2007_ill}, we always draw pairs of configurations $( \qnms^{i}, \qnms^{i+1} )$, where $\qnms^{i+1} = -\qnms^{i}$, ensuring that forces from the symmetric part of $\pot$ cancel out exactly.
    
    \item
    To compute the average Hessian we use an integral by parts, as suggested in reference \cite{brown_2013_scp}, and to further reduce the variance, we express it in terms of the difference between the harmonic and the anharmonic forces:
    \begin{equation}
    \begin{split}
        \quad \quad \expectationvalue{\hessian\left(\qats\right)}_{\ham^{\trial}_{l}}^{k} & = \hessian^{\trial}_l - \dispcorr^{-1} \expectationvalue{\left[\qats- \qats^{\trial}_l \right]^{T} \left[\fats(\qats) - \fats^{\trial}(\qats) \right] }_{\ham^{\trial}_{l}}^{k}
        \\
        \fats^{\trial}(\qats) & \equiv -\hessian^{\trial}_l \left(\qats- \qats^{\trial}_l\right) 
        \end{split}
    \end{equation}
\end{enumerate}
Samples are drawn in sets of $N_s$ until at least one component of the average forces (in terms of normal mode coordinates) is statistically significant, as assessed by whether the average over samples is larger than the standard deviation. 

We only update $\qats^{\trial}$ along those normal modes which exhibit a statistically significant net force, $\expectationvalue{\fnm{i}}$, where $\expectationvalue{\fnms} = {\tilde {\bf U}}^T \expectationvalue{\fats(\qats)}$. Direct application of Eq.~\ref{eq:scp_optimization} (b) in Cartesian space may lead to instability due to the residual statistical errors. The optimization continues until no statistically significant force component remain or the ``batch weights'' become smaller than a preset threshold, at which point a new SCP iteration begins.

While all modes must be real upon convergence, insufficient statistics may lead to spurious imaginary modes with $\omega_i^2 < 0$ before convergence is achieved. Such imaginary modes are treated by setting $\omega_i^2 = -\omega_i^2$ in the effective harmonic description.

\section{\label{sec:rnd} Results and discussion}
We first describe the systems that have been studied and the potentials that have been used to describe inter-particle interactions. We then investigate the scaling of the computational cost of the methods with respect to the system size, before assessing their accuracy by systematically comparing the approximate free energies to reference quantum thermodynamic integrations. We neglect the centre of mass contribution to the free energy throughout as it cancels out when comparing  methods. We converge all results with respect to the sampling of the vibrational BZ  by increasing the simulation cell size, allowing us to compare the different methods in equivalent and physically meaningful conditions. 

\subsection{\label{ssec:systems} Systems and Computational Details}

Three sets of materials are studied in this work in order of increasing complexity. As a first example we consider the diamond~\cite{yamanaka_1996_diamond} and lonsdaleite~\cite{bundy_1967_lonsdaleite} allotropes of carbon. These differ only in the stacking of hexagonal bilayers of tetrahedrally-coordinated carbon atoms. Their room temperature densities are identical to within experimental error \cite{yamanaka_1996_diamond,bundy_1967_lonsdaleite}, and equal to $3.51 \textrm{g}/\textrm{cm}^3$.
We consider simulation cells containing up to 64 atoms, starting from the two- and four-atom primitive cells for diamond and lonsdaleite respectively. All the cells were designed to be as close to cubic as possible to render the effective sampling of the vibrational BZ as uniform as possible. Inter-atomic interactions are modelled using the Gaussian Approximation Potential (GAP) of Deringer and Cs{\'{a}}nyi~\cite{deringer_2017_carbon_gap}, which is based on LDA DFT calculations for configurations from MD simulations of liquid and amorphous carbon. For crystalline carbon (including diamond and graphite) it has been shown to reproduce DFT energies and forces to within RMS errors of 2~meV/atom and  0.1~eV/Angstrom, respectively.
    
As a second benchmark, we discuss two proton-ordered polymorphs of ice, hexagonal (XIh)~\cite{leadbetter_1985_ice_xih,howe_1989_ice_xih} and cubic (XIc)~\cite{geiger_2014_Ic} ice. These become thermodynamically (meta-)stable below the experimental transition temperature for proton-disordering of 72\,K~\cite{leadbetter_1985_ice_xih,howe_1989_ice_xih}. For XIc we assume $I4_1/amd$ symmetry, noting that the true experimental structure of XIc is still under debate~\cite{geiger_2014_Ic}. In direct analogy to the above carbon allotropes the oxygen sublattices of XIh and XIc only differ in the stacking of bilayers of tetrahedrally-coordinated oxygen atoms. In view of the absence of experimental data for sufficiently pure XIc we take its density to be identical to that of XIh. We use the experimental density at ambient pressure and 10\,K of $0.93 \textrm{g}/\textrm{cm}^3$, noting that the thermal expansion of ice XIh between 10\,K and 70\,K is less than 0.5\,\%. We use simulation cells containing up to 16 molecules 
to allow for the possibility of coupling between pseudo-translations, which are not present at the $\Gamma$-point of the unit cell, and librational, bending and O--H bond stretching modes.  The interatomic interactions are described using a Behler-Parinello type neural network (NN)~\cite{behler_2007_nn}, based on B3LYP+D3 DFT reference calculations for around 20,000 liquid water and hexagonal ice configurations from MD and PIMD trajectories. This potential successfully reproduces the density of states, pair correlation functions and  energy fluctuations of B3LYP+D3 liquid water~\cite{kapil_2016_water_nn} and has been used to study the quantum kinetic energy, proton momentum distribution, and vibrational density of states of solid and liquid water~\cite{rossi_2018_gle4dyn, kapil_2018_water_momentum_distribution}.

Finally we analyze two polymorphs of paracetamol (N-acetyl-p-aminophenol), the monoclinic form I~\cite{haisa_1974_paracetamol_form_I} and the orthorhombic form II~\cite{haisa_1976_paracetamol_form_II}. The two forms differ in the packing of hydrogen bonded sheets of molecules --  zig-zag for form I and almost planar for form II. We consider the conventional unit cells containing four and eight formula units for forms I and II respectively, at room temperature experimental densities. Inter-atomic interactions are described on the basis of the Merk Molecular Force field also used in Ref.~\cite{rossi_2016_paracetamol}. The accuracy parameter of the PPPM method\cite{hockney}  used for calculating electrostatic interactions was set to $10^{-6}$ fractional error in the individual force components, which is smaller than the value usually required, to ensure a smooth PES. While this simple potential contains harmonic terms for bonds and angles, it remains highly anharmonic as the dihedral interaction term describes a (almost) free rotation of the methyl groups at room temperature.

\subsection{\label{subsec:computational_cost} Computational cost }
\begin{figure}
    \centering
    \includegraphics[width=\columnwidth]{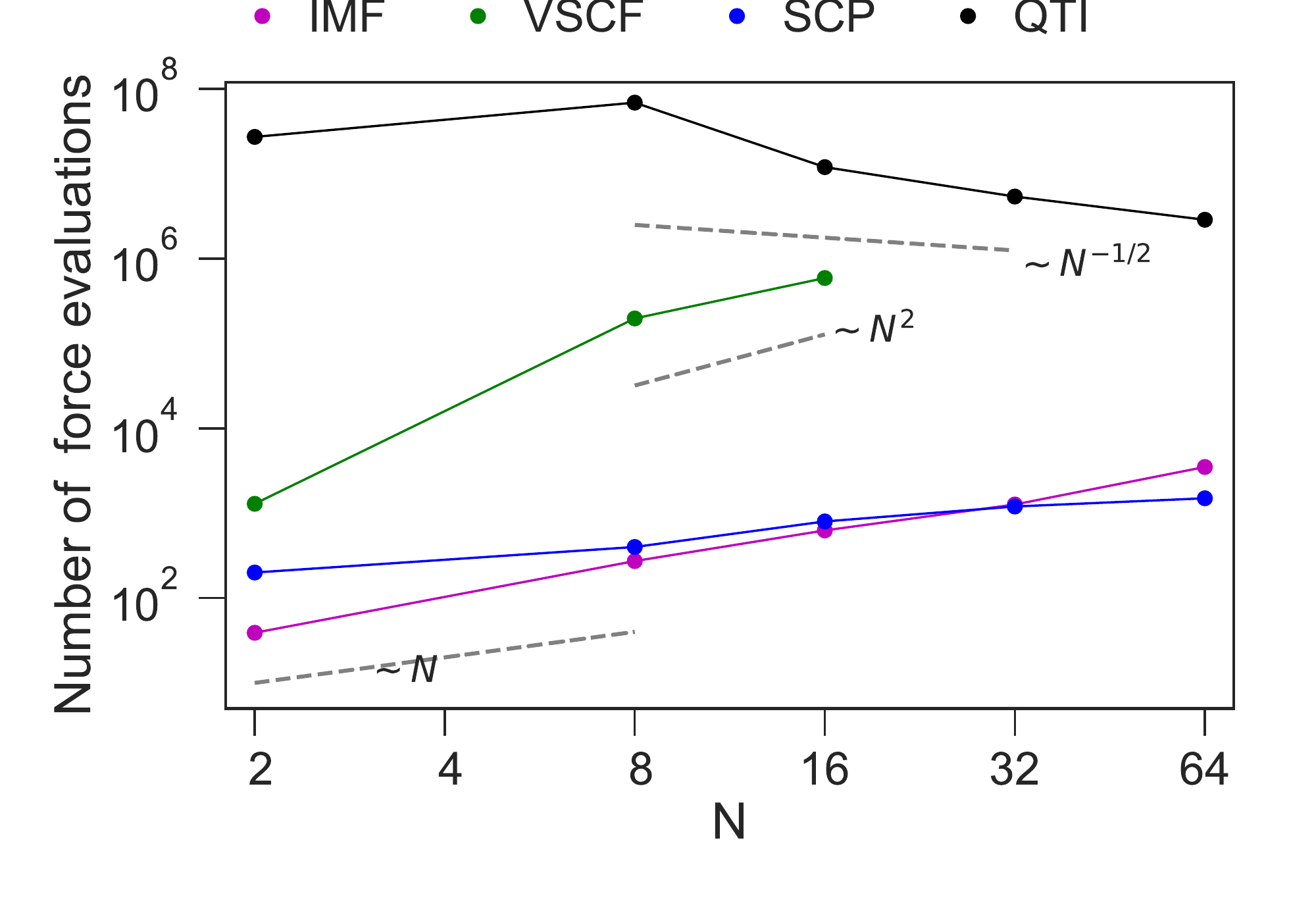}
    \caption{
Scaling of computational costs in the case of diamond in terms of the number of energy and force evaluations for IMF (pink), VSCF (green), SCP (blue) and QTI (black) with the number of atoms in simulation cell. Here, we do not make use of any crystal symmetries to reduce the cost of calculations. In all cases the free energy was converged to within 10\% of the reference TI results for 64 atoms with respect to all relevant convergence parameters.
    }
    \label{fig:costs}
\end{figure}

We define the computational cost of an approximate method ($\square$) as the minimum number of energy or force evaluations required to converge the per atom anharmomic free energy ($A^{\square} - A^{\har}$) to within 10\% of the reference value for the largest system size. For the case of diamond, this tolerance is equal to a stringent 0.2\,meV/atom. Fig.~\ref{fig:costs} shows how the cost of these methods and the reference QTI scales with the number of atoms in the simulation cell. 

The cost of IMF calculations depends linearly on the number of normal modes -- which scales linearly with system size -- and the number of points sampled along each mode. In our implementation, the later remains weakly dependent on the potential due to the variable, dynamically optimized sampling point density.

Analogously the VSCF approach exhibits a rough scaling behaviour of $N^m$ where $m$ is the dimensionality of the potential surfaces that are being sampled. For the case of diamond we use $m=2$ and therefore observe a $N^2$ dependence for large $N$. As the anharmonicity of the potential for diamond is very much dependent on BZ sampling (see SI S.1.1), the cost for the primitive cell is an outlier.

The cost for the SCP scheme, using pseudo random numbers, arises from the use of Monte Carlo importance sampling of the optimal effective harmonic description, which scales independently of system size. The statistical reweighting scheme reduces the cost for small sizes but becomes increasingly less beneficial as $N$ is increased\cite{ceriotti_2011_curse}. The net result is a near linear scaling behaviour for the system sizes that we have considered. We note that the use of Sobol sequences improves the convergence of the MC integrals for small system sizes, and thereby reduces the cost of the SCP, but leads to an unfavourable exponential scaling (see SI S1.3) for large $N$.

The reference calculations (QTI) were performed using a combination of a TI from the harmonic reference to the anharmonic potential using classical MD and a quantum TI over mass using PIMD. One should note that the variance of the integrands and the cost of performing one molecular dynamics step is different for classical and quantum MD. 
Thus the minimum number of force evaluations required to reduce the error to within the tolerance is an optimization problem detailed in the supporting information (see SI S4). Since both the integrals are effective energies and the fluctuations of the potential energy in the canonical ensemble display a $~\sim \frac{1}{\sqrt{N}}$ behaviour with respect to its mean, QTI also displays a $~\sim \frac{1}{\sqrt{N}}$ scaling behaviour.

In summary, in the limit of small system size, SCP and IMF display the most favourable scaling. The reference technique QTI displays a $\mathcal{O}(1/\sqrt{N})$ behaviour, making it  
the least expensive method in the limit of large system size. For carbon, however, as well as for all the system discussed in this work, we do not reach this limit, and QTI requires a substantially larger number of force evaluations than either SCP or IMF. It is worth noting that at fixed cell size (unlike all other methods) IMF and VSCF also provide the temperature dependence of the free energy without any additional force evaluations. %

In the current implementation, none of the above free energy methods exploit crystal symmetries. Exploiting crystal symmetries in HAR, IMF and VSCF is straightforward and the reduction in computational cost is simply related to the reduction in the number of independent normal modes. Crystal symmetries can similarly, albeit not quite as trivially, be exploited in SCP~\cite{errea_2014_scp} and the other methods. However, crystal symmetries do not affect the overall scaling behaviour with respect to the number of degrees of freedom considered in a given calculation and have therefore not been considered in the benchmarks for the computational cost. 

\subsection{\label{subsec:accuracy} Accuracy }

We gauge the accuracy of the approximate methods by studying the error incurred in the absolute anharmonic free energy, and in the free energy differences between two phases of the same material.%

\subsubsection{\label{subsubsec:carbon} Allotropes of Carbon}

\begin{figure*}
    \centering
    \includegraphics[width=\textwidth]{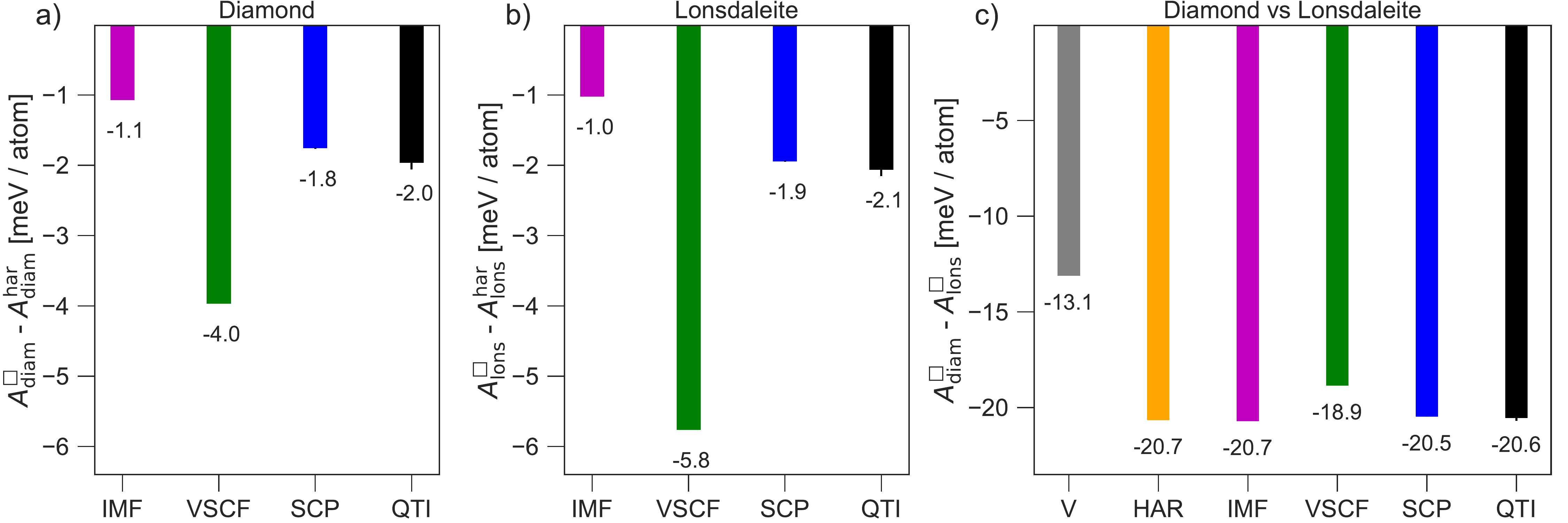}
    \caption{
    Panels (a), (b) and (c) respectively show the quantum anharmonic Helmholtz free energies $A^{\square} - A^{\har}$ of diamond and lonsdaleite allotropes of carbon, and their free energy difference $A^{\square}_{\textrm{diam}} - A^{\square}_{\textrm{lons}}$  at 300 K with IMF (pink), VSCF (green), SCP (blue), and  QTI (black).
    }
    \label{fig:carbon}
\end{figure*}

Diamond and lonsdaleite are mildly anharmonic systems which serve as excellent starting points for our study. 
We find that 32-atom simulation cells suffice to converge the free energy difference between diamond and lonsdaleite, with respect to BZ sampling. A detailed description of the workflow and system size convergence can be found in the SI (see S1.1). 

As shown in panels (a) and (b) of Fig.~\ref{fig:carbon}, the quantum anharmonic contribution to the free energy of both diamond and lonsdaleite is approximately 2\,meV/atom. IMF, which considers anharmonicity only along normal modes, underestimates the anharmonic free energy by around 1\,meV/atom. Including pairwise mean-field coupling using VSCF leads to a large over correction that increases the error to over 2\,meV/atom, while SCP (which also includes a mean-field anharmonic corrections within Gaussian statistics) gives excellent results in comparison to the reference. This indicates that the error in VSCF arises from the truncation of the potential. 

We also study the accuracy of the methods in reproducing the free energy difference between diamond and lonsdaleite, as shown in panel (c) of Fig.~\ref{fig:carbon}. 
Notably, the free energy contribution from quantum anharmonic motion for the two allotropes is almost identical so that the anharmonic free energy difference is almost identical to the harmonic free energy difference.
Fortuitously IMF and VSCF benefit from large amounts of error cancellation and reproduce the exact result within the errors in the anharmonic free energies. Overall, all approximate methods perform reasonably well at reproducing both the (very small) anharmonic corrections and the free energy difference. 

\subsubsection{\label{subsubsec:ice} Polymorphs of Ice}

\begin{figure*}
    \centering
    \includegraphics[width=\textwidth]{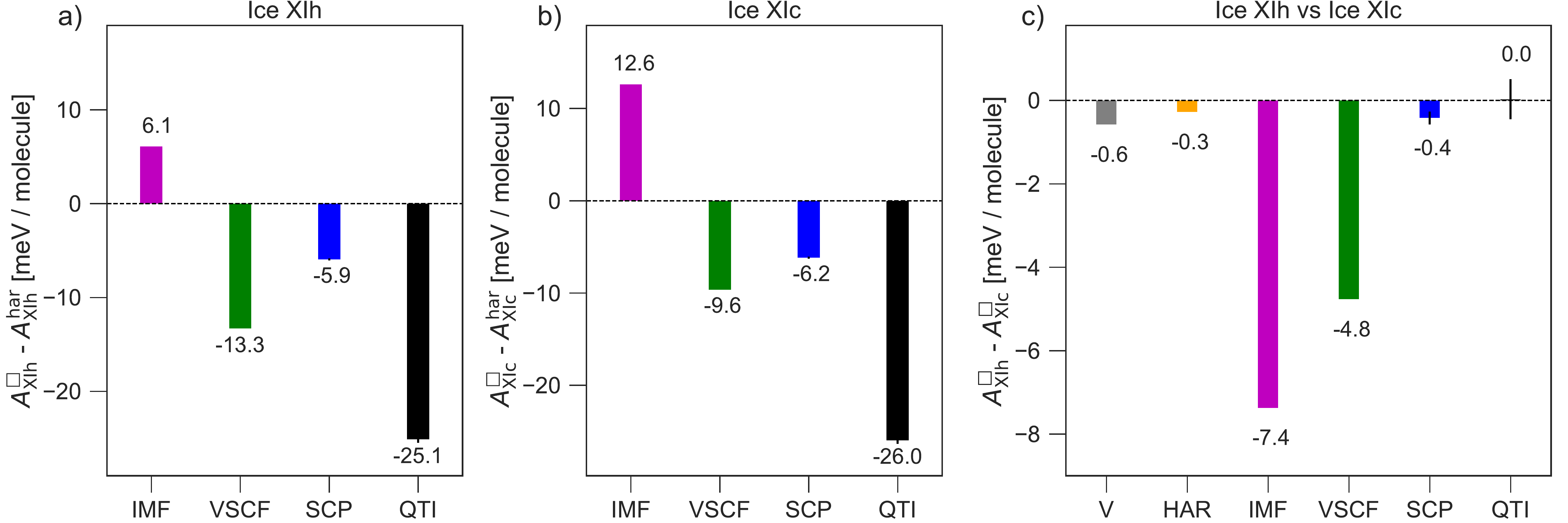}
    \caption{
    Panels (a), (b) and (c) respectively show the quantum anharmonic Helmholtz free energies $A^{\square} - A^{\har}$ of hexagonal and cubic polymorphs of ice XI, and their free energy difference $A^{\square}_{\textrm{XIh}} - A^{\square}_{\textrm{XIc}}$  at 70 K with IMF (pink), VSCF (green), SCP (blue) and  QTI (black).
    }
    \label{fig:ice}
\end{figure*}

Ice XIh and ice XIc are a more challenging test-case because of the simultaneous large anharmonic free energy due to the pronounced anharmonicity of the O--H bond, the coupling between the high and low frequency modes\cite{monacelli_2018_scp}, and the small free energy difference between the hexagonal and cubic polymorphs~\cite{engel_2015_ice, cheng_2019_ice}. 
Supercells containing 16 molecules of water suffice to converge the free energy difference for all methods. Details of the calculations can be found in the SI (see S2.1)

As shown in panels (a) and (b) of Fig.~\ref{fig:ice}, the overall contribution from quantum anharmonicity to the free energy is around 25\,meV/molecule for both systems. Contrary to the case of carbon, we find the approximate methods do not accurately reproduce the reference anharmonic free energy. For instance, the IMF technique produces qualitatively incorrect anharmonic corrections. The VSCF approach with pairwise couplings of normal modes provides the best approximation, but remains off by over 10\,mev/molecule. The SCP scheme incurs errors of around 20\,meV/molecule. %

In line with previous path integral calculations on hexagonal and cubic ice\cite{cheng_2019_ice}, we find the free energy difference between the polymorphs of ice XI to be almost zero, as shown in panel (c) of Fig.~\ref{fig:ice}. IMF predicts the hexagonal form to be more stable by around 7\,meV/molecule. After adding mean field coupling corrections within VSCF the margin of stability reduces to around 5\,meV/molecule. The SCP scheme benefits from cancellation of errors and fortuitously gives the correct result within 1\,mev/molecule.%
\subsubsection{\label{subsubsec:paracetamol} Polymorphs of Paracetamol }

As a final test, we consider forms 1 and 2 of crystalline paracetamol. These are more complex molecular crystals, for which free energy calculations are complicated by the presence of quasi-free rotations of the methyl groups. 
Reference free energies are obtained by classical TI with respect to temperature~\cite{moustafa_2015_ti} from 10\,K to 300\,K followed by quantum TI over masses. To ensure that all of the three degenerate rotational conformers of the paracetamol molecules, are explored evenly at low temperatures, the classical TI is performed using replica exchange molecular dynamics~\cite{sugita_1999_remd}. 
At 10\,K classical anharmonicity is almost completely suppressed and the classical anharmonic free energy can be accurately estimated using a harmonic approximation (accounting for the three degenerate conformers). For reference, the free energies were recalculated using the TI route employed in Ref.~\cite{rossi_2016_paracetamol} which proves to reproduce the same result. 

\begin{figure*}
    \centering
    \includegraphics[width=\textwidth]{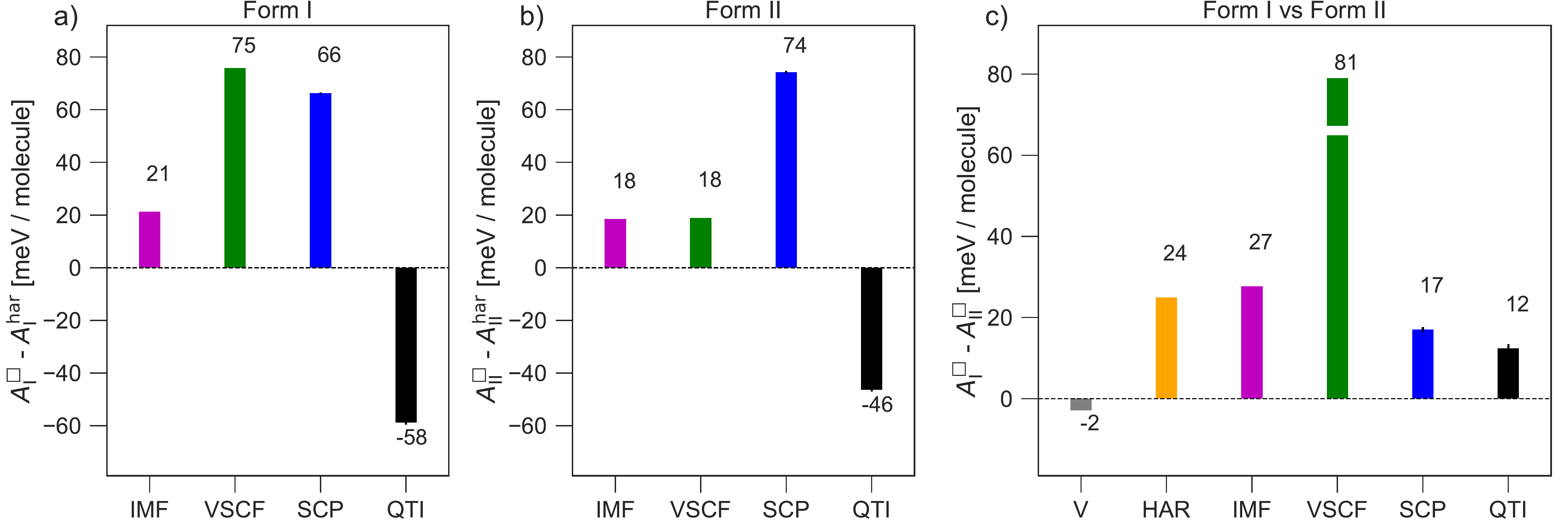}
    \caption{
Panels (a), (b) and (c) respectively show the quantum anharmonic Helmholtz free energies $A^{\square} - A^{\har}$ of form I and form II polymorphs of crystalline paracetamol, and their free energy difference $A^{\square}_{\textrm{I}} - A^{\square}_{\textrm{II}}$  at 300 K with IMF (pink), VSCF (green), SCP (blue) and  QTI (black).
}
    \label{fig:paracetamol}
\end{figure*}

As shown in Figs.~\ref{fig:paracetamol} (a) and (b), the overall quantum anharmonic corrections for forms I and II are around -58 and -46\,meV/molecule. The degeneracy of the conformers contributes the dominant part, $-k_B T \ln(3) \approx -28$\,meV / molecule. 
All approximate anharmonic methods produce qualitatively incorrect free energy corrections due to anharmonicity. Furthermore, just as the harmonic approximation, they are unable to account for the degeneracy of the three conformers, that would have to be identified and corrected for manually. 

As shown in Fig.~\ref{fig:paracetamol} (c), the free energy difference between the two forms is around 12\,meV/molecule. The difference with respect to Ref.\cite{rossi_2016_paracetamol} arises due to the use of slightly different lattice constants (see SI S3.1.4), a more accurate path integral sampling technique\cite{kapil_2016_water_nn} and a 
finer PPPM mesh for the Ewald summation of electrostatics.
As in the cases of carbon and ice, the IMF and SCP benefit from significant error cancellation. The former correctly predicts form II to be more stable but performs worse than a harmonic approximation in getting the correct magnitude. The latter also predicts the correct sign but fortuitously estimates the magnitude to within 5 mev/molecule of the exact result. VSCF doesn't benefit from error cancellation to the same extent and overestimates the stability of form I by over 80\,meV/molecule.

The failure of normal mode based approaches for paracetamol is unsurprising, as the description of quasi-free rotations requires curvilinear coordinates. 
In paracetamol the potential energy barrier for rotational motion correspond to approximately 200\,K$k_B$ implies (even classically) quasi-free rotation of the methyl groups at room temperature.
For the force field used to describe paracetamol, the potential governing rotation and breathing of methyl groups can be extracted explicitly (neglecting coupling to the remainder of the molecule) and takes the simple form
\begin{equation}
    V(r,\theta) = \frac{1}{2} k \left( r - r_0 \right)^2 + V_\theta \left(1 - \cos{(3\theta)}\right)
\end{equation}
where 
$k = 53.114$\,eV/\AA$^2$, $r_0 = 1.09$\,\AA, and $V_\theta = 8.54$\,meV. This allows us to study the failure of the above approximate free energy methods with the activation of angular motion on the basis of a simple toy model, 
\begin{equation}
    \ham = - \frac{\hbar^2}{2 \mu} \left[ \frac{\partial^2}{\partial r^2} + \frac{1}{r^2} \frac{\partial^2}{\partial \theta^2} \right] + V(r,\theta)
\end{equation}
which can easily be studied over a range of temperatures. The exact solution for this simple model is obtained by exact diagonalization (ED) of the Hamiltonian matrix on a regular, square two dimensional real space grid of $256 \times 256$ points spanning from $(x,y) = (-1.5 r_0, -1.5 r_0)$ to $(x,y) = (1.5 r_0, 1.5 r_0)$. We find that the reference free energy is converged to within 0.2\,meV.

\begin{figure}
    \centering
    \includegraphics[width=\columnwidth]{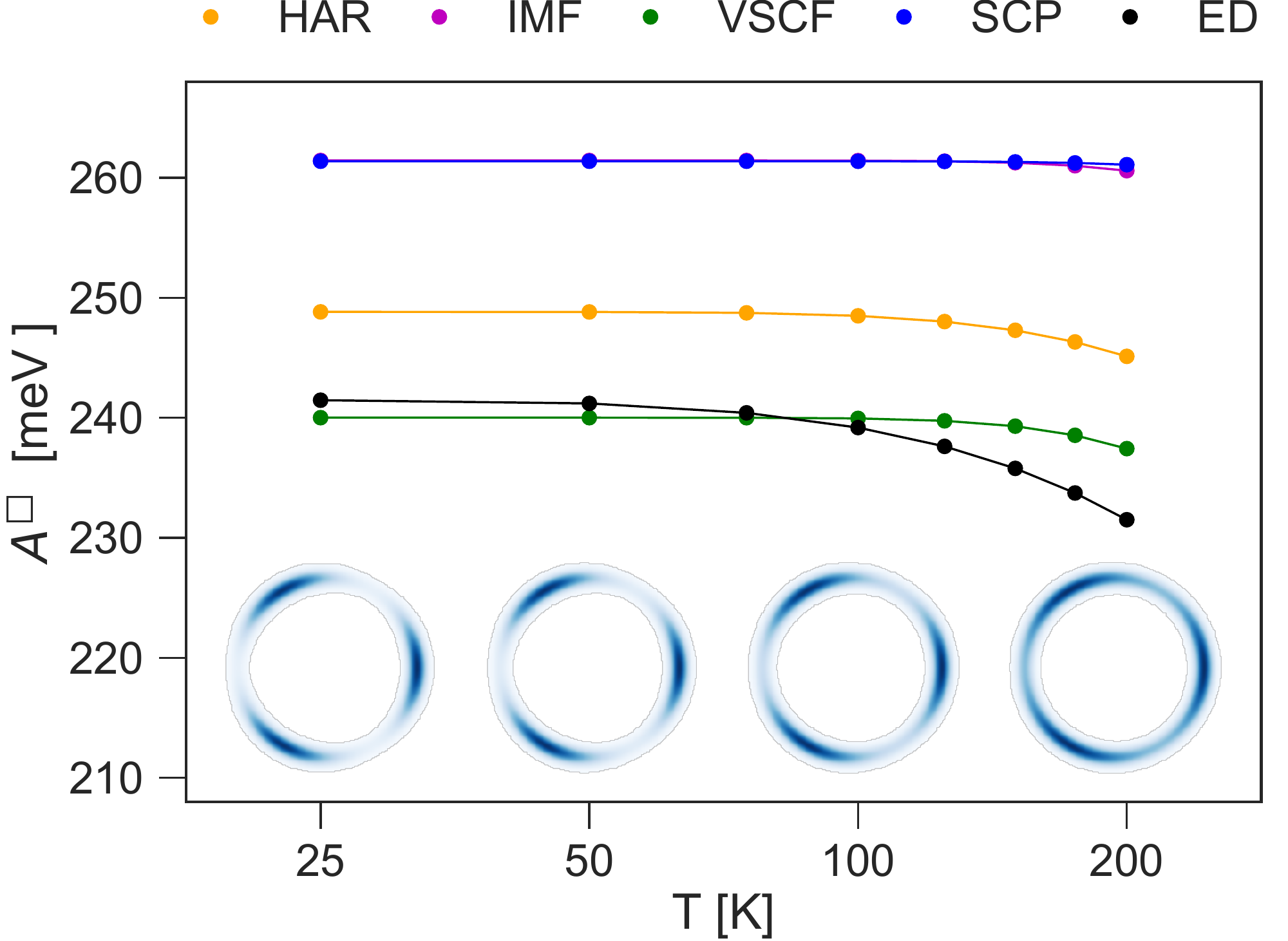}
    \caption{Temperature dependence of the Helmholtz free energy of a quasi-free rotor obtained with IMF (pink), VSCF (green), SCP (blue). Reference data obtained from exact diagonalization are shown in black. The insets show the position distribution function at temperatures of 25\,K, 50\,K, 100\,K and 200\,K with blue and white indication high and low probability.}
    \label{fig:paracetamol_model}
\end{figure}

The temperature dependence of the free energy of the model system is shown in Fig.~\ref{fig:paracetamol_model}. Since none of the approximate methods are able to account for the three fold degeneracy of the system,  we remove $-k_B T \ln{3}$ from the exact results and study the anharmonic free energy of one of the minima. 
The results show that above around 75\,K the VSCF approximation becomes increasingly inaccurate as the amplitude of angular motion of methyl groups increases and the vibrational density delocalizes over the three equivalent potential energy minima (see Fig.~\ref{fig:paracetamol_model}). 
Furthermore, the harmonic, IMF, and SCP approximations severely overestimate of the free energy throughout. For the harmonic and IMF approximation this can simply be explained by the fact that linear coordinates mix angular and much higher frequency radial motion, so that the effective ``angular mode'' is stiffened substantially, while the radial mode retains the true harmonic frequency. 

\section{\label{sec:conclusions}Conclusions}

Diamond and lonsdaleite, as examples of simple quasi-harmonic solids, highlight the utility of approximate free energy methods. While the accuracy of the approximate Helmholtz free energies varies, all approaches achieve sub-2\,meV/atom accuracy and, more importantly, consistently yield a systematic improvement over the harmonic approximation at a substantially lower computational cost than the reference QTI. 
On the other hand, ice and paracetamol, as examples of  more complex, molecular crystals, highlight the limitations of approximate techniques. The free energies of the molecular crystals are substantially overestimated due to the inherent limitations of normal modes based descriptions in the presence of large-amplitude curvilinear librational or quasi-free rotational motion. The simple model description of the rotation of the methyl group in paracetamol demonstrates that SCP, IMF and VSCF artificially stiffen the rotational modes, leading to an overestimation of the total free energy. The failure of these methods for ice can also be understood along the same lines: at larger displacements along the normal modes initially corresponding to librational motion, O--H bonds are stretched and bent, leading to an overestimation of the effective frequency of librational motion and consequently the free energy. This is confirmed by the blue shifts of the librational modes observed in the case of IMF and SCP with respect to the harmonic approximation (see SI S2.3). Consequently, these methods do not consistently yield systematic improvements over the harmonic approximation for free energies of solids that possess high amplitude librational or quasi-rotational modes. However, we expect these techniques to perform well for atomic and ionic solids, where the point-like nature of the atomic/ionic building blocks suppresses large-amplitude curvilinear motion. %
It is worth mentioning that (in suitable applications) normal modes based approximate methods lend themselves to identifying the atomistic/structural origins of anharmonicity and facilitate analyses of, for example, spectral properties of strongly anharmonic phonons, as probed by inelastic scattering processes, the formation charge-density-waves, and ferroelectric instabilities~\cite{bianco_2017_scp}.

We also find that the approximate methods methods benefit from error cancellation, leading to errors in free energy differences that are consistently smaller than the errors in the absolute Helmholtz free energies. However, it is worth noting that we have compared systems with very similar local environments. In general such beneficial cancellation of errors is not guaranteed. We demonstrate this in section S5 of the SI by studying the free energy of a few high density phases of ice (II, IX and XV) relative to that of XIh. As shown in in Fig.~S5, SCP -- that benefits from error cancellation when comparing the cubic and hexagonal forms of ice -- does not yield a qualitatively correct order of relative free energies. 

As the approximate results can vary from almost quantitatively accurate to qualitatively incorrect results, QTI is the only free energy methods among those considered in this work that provides reliable anharmonic free energies for large and complex organic solids. Given that it displays a $\mathcal{O}(1/\sqrt{N})$ computational cost, it may furthermore require comparable or fewer force evaluations than a SCP or VSCF calculation for systems of interest, in particular when considering biological or pharmaceutical compounds that involve large unit cells with flexible molecular units. 
It is further worth noting that QTI (and other statistical sampling methods) are substantially less susceptible to noise in the underlying PES than the harmonic approximation, IMF and VSCF. While random noise largely cancels out in the ensemble averages calculated in statistical sampling methods, especially in combination with stochastic thermostats~\cite{Khne_2007_noisy_langevin, Mandal_2018_noisy_langevin}, the harmonic approximation relies on the ability to determine a meaningful dynamical matrix and thus a differentiable PES, and the IMF and VSCF require an interpolatable PES. 
This is demonstrated by performing the Ewald summation in the description of paracetamol using a coarser PPPM mesh, which leads to discontinuities in the PES of 0.50 meV (see SI S6). 
While the resultant QTI and SCP free energies remain largely unaffected, the free energy estimates obtained from the analytic methods -- at least in the  implementation we discuss here -- depend heavily on the size of the finite displacements underlying the mapping of the PES and the VSCF in particular eventually fails to converge altogether.

Efforts towards obtaining a computationally feasible anharmonic free energy should therefore be channelled towards reducing the cost of performing a QTI or, at least, a classical-nuclei TI with nuclear quantum contributions evaluated at a more approximate level \cite{rossi_2016_paracetamol,kapil_2019_mofs}. This includes streamlining hierarchical frameworks\cite{ duff_2015_uptild_2} that perform the full free energy calculations using inexpensive bespoke potentials\cite{hellman_2013_tdep} or cheaper basis sets\cite{grabowski_2009_al}, reducing the cost to that of reversibly switching an \emph{ab initio} potential.
Machine learning potentials offer exceptional promise to provide ab-initio-quality potential energy surfaces to evaluate the anharmonic free energy~\cite{cheng_2019_ice}, and approximate methods could also constitute an effective sampling approach to generate data to train and validate such ML potentials.  
\section{Acknowledgements}
VK and EAE contributed equally to the work. We would like to thank Francesco Mauri, Marco Cherubini for helping us validate our SCP implementation. Bartomeu Monserrat, is also acknowledged for providing a critical reading of an early version of the manuscript.  MC and EAE, acknowledge funding by the European Research Council under the European Union's Horizon 2020 research and innovation program (grant no. 677013-HBMAP). MR acknowledges funding from BiGmax, the Max Planck research network on big-data-driven materials science.

\section{Supporting Information}

Details of calculations on allotropes of carbon, ice XI and paracetamol; Analysis of efficiency of the thermodynamic integration route; Relative stability of some of the high density proton ordered phases of ice; Influence of noise in the potential energy landscapes on IMF
\providecommand{\latin}[1]{#1}
\makeatletter
\providecommand{\doi}
  {\begingroup\let\do\@makeother\dospecials
  \catcode`\{=1 \catcode`\}=2 \doi@aux}
\providecommand{\doi@aux}[1]{\endgroup\texttt{#1}}
\makeatother
\providecommand*\mcitethebibliography{\thebibliography}
\csname @ifundefined\endcsname{endmcitethebibliography}
  {\let\endmcitethebibliography\endthebibliography}{}

\include{si-pdf}

\end{document}